% Luca Bombelli, Statistical Lorentzian geometry..., gr-qc/0002053
% ------------------------- 78 characters per line ------------------------- %
% general settings
\magnification=\magstep1\raggedbottom\def\normalbaselineskip{13pt}
\fontdimen16\tensy=2.7pt\fontdimen17\tensy=2.7pt
\abovedisplayskip=6pt plus 3pt minus 3pt
\belowdisplayskip=6pt plus 3pt minus 3pt
% general formatting and math macros
\def\section#1#2{\goodbreak\vskip0.6cm\leftskip=20pt\parindent=-20pt\indent
{\bf\hbox to 20pt{#1\hss}#2}\medskip\nobreak\leftskip=0pt\parindent=20pt}
\def\C{{\cal C}}\def\sss#1{{\scriptscriptstyle{#1}}}
\def\g{{\bf g}}\def\rtg{\ts{\sqrt{-g}}}\def\tw{\widetilde}\def\ee{{\rm e}}
\def\dd{{\rm d}}\def\DD{{\sss D}}\def\ts#1{{\textstyle{#1}}}
\def\lp{\ell_{\sss{\rm P}}}\def\NN{{\rm I\hskip-2pt N}}
\def\real{{\rm I\hskip-2pt R}}\def\ovr{\overline}\def\nsum{\sum\nolimits}
\def\half{\ts{1\over2}}\def\cut{\hfil\break}
% Extra fonts (may need to change the fonts cm??8 to am??8 sometimes)

% symbols for special posets
\def\uno{\raise1.5pt\hbox{$\sss\bullet$}}
\def\conn{\kern3pt\raise0.4pt\hbox{\vrule width.3pt height4.5pt}%
\kern-1.7pt\lower2pt\hbox{$\sss\bullet$}%
\kern-3.7pt\raise5pt\hbox{$\sss\bullet$}\kern1pt}
\def\disc{\raise1.5pt\hbox{$\sss{\bullet\;\bullet}$}}
\def\tre{\raise1.5pt\hbox{$\sss{\bullet\;\bullet\;\bullet}$}}
\def\due{\raise1.5pt\hbox{$\sss\bullet$}\conn}
\def\eev{\kern2pt\hbox{$\land%
\kern-7.6pt\lower2pt\hbox{$\sss\bullet$}%
\kern-1.1pt\raise5pt\hbox{$\sss\bullet$}%
\kern-1.1pt\lower2pt\hbox{$\sss\bullet$}$}\kern1pt}
\def\vee{\kern2pt\hbox{$\lor%
\kern-7.6pt\raise5pt\hbox{$\sss\bullet$}%
\kern-1.1pt\lower2pt\hbox{$\sss\bullet$}%
\kern-1.1pt\raise5pt\hbox{$\sss\bullet$}$}\kern1pt}
\def\lin{\kern3pt\raise-0.5pt\hbox{\vrule width.3pt height8pt}%
\kern-1.7pt\lower3pt\hbox{$\sss\bullet$}%
\kern-3.7pt\raise2pt\hbox{$\sss\bullet$}%
\kern-3.7pt\raise7pt\hbox{$\sss\bullet$}\kern1pt}
% Second part of the macros, based on a shortened version of jnl
\overfullrule=0pt
\def\normalparindent{24pt}
\def\beginparmode{\endmode\begingroup \def\endmode{\par\endgroup}}
\let\endmode=\par
\def\\{\cr}
\def\body{\beginparmode\parindent=\normalparindent}
\def\head#1{\par\goodbreak{\immediate\write16{#1}
    {\noindent\bf #1}\par}\nobreak\nobreak}
\def\refto#1{$[#1]$}\def\ref#1{Ref.~#1}\def\refs#1{Refs.~$#1$}
\def\Ref#1{Ref.~$#1$}\def\cite#1{{#1}}
\def\(#1){(\call{#1})}
\def\call#1{{#1}}\def\taghead#1{{#1}}
\def\references{\head{REFERENCES}\beginparmode\frenchspacing\parskip=0pt}
\gdef\refis#1{\item{#1.\ }}
\gdef\journal#1,#2,#3,#4.{{\sl #1}\ {\bf #2} (#3) #4} % new def
\gdef\Journal#1,#2,#3,#4.{#1~{\bf #2}, #3 (#4)}
\def\endreferences{\body}
\def\endit{\endmode\vfill\supereject}\let\endpaper=\endit
%
% REFORDER
\catcode`@=11
\newcount\r@fcount \r@fcount=0\newcount\r@fcurr
\immediate\newwrite\reffile\newif\ifr@ffile\r@ffilefalse
\def\w@rnwrite#1{\ifr@ffile\immediate\write\reffile{#1}\fi\message{#1}}
\def\writer@f#1>>{}
\def\citeall#1{\xdef#1##1{#1{\noexpand\cite{##1}}}}
\def\cite#1{\each@rg\citer@nge{#1}}
\def\each@rg#1#2{{\let\thecsname=#1\expandafter\first@rg#2,\end,}}
\def\first@rg#1,{\thecsname{#1}\apply@rg}
\def\apply@rg#1,{\ifx\end#1\let\next=\relax%
    \else,\thecsname{#1}\let\next=\apply@rg\fi\next}%
\def\citer@nge#1{\citedor@nge#1-\end-}
\def\citer@ngeat#1\end-{#1}
\def\citedor@nge#1-#2-{\ifx\end#2\r@featspace#1
    \else\citel@@p{#1}{#2}\citer@ngeat\fi}
\def\citel@@p#1#2{\ifnum#1>#2{\errmessage{Reference range #1-#2\space is bad.}
    \errhelp{If you cite a series of references by the notation M-N, then M
    and N must be integers, and N must be greater than or equal to M.}}\else%
    {\count0=#1\count1=#2\advance\count1 by1\relax\expandafter\r@fcite\the%
    \count0,\loop\advance\count0 by1\relax %	Loop from M to N
    \ifnum\count0<\count1,\expandafter\r@fcite\the\count0,%
    \repeat}\fi}
\def\r@featspace#1#2 {\r@fcite#1#2,}
\def\r@fcite#1,{\ifuncit@d{#1}\expandafter\gdef\csname r@ftext\number%
    \r@fcount\endcsname {\message{Reference #1 to be supplied.}%
    \writer@f#1>>#1 to be supplied.\par }\fi\csname r@fnum#1\endcsname}
\def\ifuncit@d#1{\expandafter\ifx\csname r@fnum#1\endcsname\relax%
    \global\advance\r@fcount by1%
    \expandafter\xdef\csname r@fnum#1\endcsname{\number\r@fcount}}
\let\r@fis=\refis
\def\refis#1#2#3\par{\ifuncit@d{#1}%
    \w@rnwrite{Reference #1=\number\r@fcount\space is not cited up to now.}%
    \fi\expandafter\gdef\csname r@ftext\csname r@fnum#1\endcsname\endcsname%
    {\writer@f#1>>#2#3\par}}
\def\r@ferr{\endreferences\errmessage{I was expecting to see
    \noexpand\endreferences before now; I have inserted it here.}}
\let\r@ferences=\references
\def\references{\r@ferences\def\endmode{\r@ferr\par\endgroup}}
\let\endr@ferences=\endreferences
\def\endreferences{\r@fcurr=0{\loop\ifnum\r@fcurr<\r@fcount
    \advance\r@fcurr by 1\relax\expandafter\r@fis\expandafter{\number\r@fcurr}%
    \csname r@ftext\number\r@fcurr\endcsname%
    \repeat}\gdef\r@ferr{}\endr@ferences}
\let\r@fend=\endpaper
\gdef\endpaper{\ifr@ffile\immediate\write16{Cross References
    written on []\jobname.REF.}\fi\r@fend}
\catcode`@=12
\citeall\refto\citeall\ref\citeall\Ref\citeall\refs
%
% EQNORDER
\catcode`@=11
\newcount\tagnumber\tagnumber=0
\immediate\newwrite\eqnfile\newif\if@qnfile\@qnfilefalse
\def\write@qn#1{}\def\writenew@qn#1{}
\def\w@rnwrite#1{\write@qn{#1}\message{#1}}
\def\@rrwrite#1{\write@qn{#1}\errmessage{#1}}
\def\taghead#1{\gdef\t@ghead{#1}\global\tagnumber=0}
\def\t@ghead{}
\expandafter\def\csname @qnnum-3\endcsname
    {{\t@ghead\advance\tagnumber by -3\relax\number\tagnumber}}
\expandafter\def\csname @qnnum-2\endcsname
    {{\t@ghead\advance\tagnumber by -2\relax\number\tagnumber}}
\expandafter\def\csname @qnnum-1\endcsname
    {{\t@ghead\advance\tagnumber by -1\relax\number\tagnumber}}
\expandafter\def\csname @qnnum0\endcsname
    {\t@ghead\number\tagnumber}
\expandafter\def\csname @qnnum+1\endcsname
    {{\t@ghead\advance\tagnumber by 1\relax\number\tagnumber}}
\expandafter\def\csname @qnnum+2\endcsname
    {{\t@ghead\advance\tagnumber by 2\relax\number\tagnumber}}
\expandafter\def\csname @qnnum+3\endcsname
    {{\t@ghead\advance\tagnumber by 3\relax\number\tagnumber}}
\def\equationfile{\@qnfiletrue\immediate\openout\eqnfile=\jobname.eqn%
    \def\write@qn##1{\if@qnfile\immediate\write\eqnfile{##1}\fi}
    \def\writenew@qn##1{\if@qnfile\immediate\write\eqnfile
    {\noexpand\tag{##1} = (\t@ghead\number\tagnumber)}\fi}}
\def\callall#1{\xdef#1##1{#1{\noexpand\call{##1}}}}
\def\call#1{\each@rg\callr@nge{#1}}
\def\each@rg#1#2{{\let\thecsname=#1\expandafter\first@rg#2,\end,}}
\def\first@rg#1,{\thecsname{#1}\apply@rg}
\def\apply@rg#1,{\ifx\end#1\let\next=\relax%
    \else,\thecsname{#1}\let\next=\apply@rg\fi\next}
\def\callr@nge#1{\calldor@nge#1-\end-}
\def\callr@ngeat#1\end-{#1}
\def\calldor@nge#1-#2-{\ifx\end#2\@qneatspace#1 %
    \else\calll@@p{#1}{#2}\callr@ngeat\fi}
\def\calll@@p#1#2{\ifnum#1>#2{\@rrwrite{Equation range #1-#2\space is bad.}
    \errhelp{If you call a series of equations by the notation M-N, then M
    and N must be integers, and N must be greater than or equal to M.}}\else%
    {\count0=#1\count1=#2\advance\count1 by1\relax\expandafter\@qncall\the%
    \count0, \loop\advance\count0 by1\relax%
    \ifnum\count0<\count1,\expandafter\@qncall\the\count0, \repeat}\fi}
\def\@qneatspace#1#2 {\@qncall#1#2,}
\def\@qncall#1,{\ifunc@lled{#1}{\def\next{#1}\ifx\next\empty\else
    \w@rnwrite{Equation number \noexpand\(>>#1<<) has not been defined yet.}
    >>#1<<\fi}\else\csname @qnnum#1\endcsname\fi}
\let\eqnono=\eqno
\def\eqno(#1){\tag#1}
\def\tag#1$${\eqnono(\displayt@g#1 )$$}
\def\aligntag#1\endaligntag $${\gdef\tag##1\\{&(##1 )\cr}\eqalignno{#1\\}$$
    \gdef\tag##1$${\eqnono(\displayt@g##1 )$$}}

\def\eqalignno#1{\displ@y \tabskip\centering
    \halign to\displaywidth{\hfil$\displaystyle{##}$\tabskip\z@skip
    &$\displaystyle{{}##}$\hfil\tabskip\centering
    &\llap{$\displayt@gpar##$}\tabskip\z@skip\crcr #1\crcr}}
\def\displayt@gpar(#1){(\displayt@g#1 )}
\def\displayt@g#1 {\rm\ifunc@lled{#1}\global\advance\tagnumber by1
    {\def\next{#1}\ifx\next\empty\else\expandafter
    \xdef\csname @qnnum#1\endcsname{\t@ghead\number\tagnumber}\fi}%
    \writenew@qn{#1}\t@ghead\number\tagnumber\else
    {\edef\next{\t@ghead\number\tagnumber}%
    \expandafter\ifx\csname @qnnum#1\endcsname\next\else%
    \w@rnwrite{Equation \noexpand\tag{#1} is a duplicate number.}\fi}%
    \csname @qnnum#1\endcsname\fi}
\def\ifunc@lled#1{\expandafter\ifx\csname @qnnum#1\endcsname\relax}
\let\@qnend=\end
\gdef\end{\if@qnfile\immediate\write16{Equation numbers
    written on []\jobname.EQN.}\fi\@qnend}
\catcode`@=12
% ------------------------- 78 characters per line ------------------------- %
\line{\hfill gr-qc/0002053}
\line{\hfill revised June 2000}
\vfill
\centerline{\bf Statistical Lorentzian geometry}
\centerline{\bf and the closeness of Lorentzian manifolds}
\vskip1cm
\centerline{Luca Bombelli}
\smallskip
\centerline{\sl Department of Physics and Astronomy}
\centerline{\sl 108 Lewis Hall, University of Mississippi}
\centerline{\sl USA - Oxford, MS 38677}
\centerline{E-mail: luca@phy.olemiss.edu}
\vfill
\centerline{Abstract}
\midinsert\narrower
\noindent I introduce a family of closeness functions between causal
Lorentzian geometries of finite volume and arbitrary underlying topology.
When points are randomly scattered in a Lorentzian manifold, with uniform
density according to the volume element, some information on the topology and
metric is encoded in the partial order that the causal structure induces
among those points; one can then define closeness between Lorentzian
geometries by comparing the sets of probabilities they give for obtaining the
same posets. If the density of points is finite, one gets a pseudo-distance,
which only compares the manifolds down to a finite volume scale, as
illustrated here by a fully worked out example of two 2-dimensional manifolds
of different topology; if the density is allowed to become infinite, a true
distance can be defined on the space of all Lorentzian geometries. The
introductory and concluding sections include some remarks on the motivation
for this definition and its applications to quantum gravity.
\endinsert
\baselineskip=\normalbaselineskip
\centerline{PACS numbers 04.20.Gz, 02.40.-k}
\vfill
\noindent Running head: Statistical Lorentzian geometry\hfill
\eject

\section{I.}{Introduction}

\noindent The purpose of this paper is to propose a definition of closeness
between Lorentzian geometries, where by Lorentzian geometry I mean a
diffeomorphism equivalence class $G = \{(M,\g)\}$ of manifolds with Lorentzian
metrics. More specifically, I will first define a pseudo-distance function
$d_n(G,G')$ of two geometries $G = \{(M,\g)\}$ and $G' = \{(M',\g')\}$ with
finite volumes $V_M$ and $V_{M'}$, depending on an integer $n$, such that
whenever $d_n(G,G')$ is small, the two geometries are close at large volume
scales compared to $V_M/n$ and $V_{M'}/n$, up to a global scale
transformation; most of the paper is devoted to this pseudo-distance and its
properties, but I will also extend the definition to a distance function
$d_\ell(G,G')$ depending on a length parameter $\ell$. Notice that the
geometries in question can be based on two entirely different manifolds
$M$ and $M'$.

There are various contexts in which such a definition is useful, but the ones
that motivated this work are mostly related to quantum gravity. There is a
growing amount of evidence, initially suggested by analogies with other
theories and simple consistency arguments but increasingly supported by more
rigorous results, that the structure of spacetime at the smallest scales (of
the order of or smaller than the Planck volume $\lp^4 = (G\hbar/c^3)^2$
---just for this equation, $G$ stands for Newton's gravitational constant)
differs significantly from that of the four-dimensional, topologically flat
differentiable manifold we use as a model in ordinary physics
\refto{Gibb, Kemp, Asht}. Very many different proposals exist for what to
replace this manifold with; I will mention only a few of them here, as
examples of situations in which one needs to talk about the closeness of
Lorentzian geometries.

If one assumes that large quantum fluctuations of the metric on small scales
will be associated with fluctuations in the topology itself, but
differentiable manifolds are still valid models for the geometry, one is led
to the notion of {\it spacetime foam\/} \refto{Whee,Hawk,CrSm}, a bubbling
topological magma in which topological entities like geons and wormholes
fluctuate into and out of existence. Spacetime is a quantum superposition of
differentiable manifolds of different topology, and the ones that contribute
most to the classical spacetime we see are such that each topological
fluctuation occupies on the average one Planck volume; at larger scales they
are all thought to be close to each other, and essentially indistinguishable
from a topologically flat manifold.

On the other hand, there are hints that the very notion of manifolds and
continuity may have to be abandoned for models that describe spacetime at
Planck scales. In the {\it causal set\/} proposal, spacetime is considered as
a locally finite partially ordered set \refto{Myrh, Hoof, BLMS, Bomb, Sork,
BrGr}; if the elements are thought of as events, occupying on the average one
Planck volume each, the partial order is interpreted as giving the causal
relations between them. In {\it spin foam\/} type proposals, the basic
structure is also that of a graph, but with extra variables attached to the
edges and vertices \refto{ReRo, Baez, DePi, Iwas, MaSm, Gupt}. In either
case, the continuum and the rest of the Lorentzian manifold structure we see
at large scales emerge as a thermodynamic limit, much like the description of
a gas by thermodynamic quantities such as pressure and temperature emerges at
large scales. Part of the reason why this limit exists is that, even though
there are infinitely many Lorentzian manifolds which can smoothly interpolate
between the elements of a given discrete set, they are all supposedly
indistinguishable at scales larger than $\lp^4$.

A definition of closeness between Lorentzian metrics on the same manifold $M$,
in the form of a scale dependent function $d_\lambda(\g,\g')$, has already
been given in Ref \cite{BoSo} (the main idea can also be found in Ref
\cite{BoMe}). However, that definition is not diffeomorphism invariant,
in the sense that, if $\phi$ is a diffeomorphism of $M$, in general
$d_\lambda(\g,\g') \ne d_\lambda(\g, \phi^\ast\g')$. In principle, given
such a $d_\lambda$ one can construct an invariant one \refto{BoSo}, but in
this case $d_\lambda$ is difficult to work with; and, more importantly, it is
not defined for metrics on different manifolds. My goal here is to set up a
definition that is applicable to any two Lorentzian manifolds, analogously to
the one given for Riemannian geometries by Gromov using geometrical concepts
\refto{Grom} or by Seriu using spectral techniques \refto{Seri}.
Unfortunately, the ideas behind those distances rely heavily on the
positive-definite nature of the metrics; the one I use here comes instead
from causal set theory: $G$ and $G'$ are close if, when we distribute the
same number of points at random with uniform density in (one representative
of) each of them, the probability of obtaining any given induced partial
order among those points is about the same in the two cases. A few of the
ideas that led to this work appeared earlier in a different form in Ref
\cite{BSal}.

The use of uniform distributions of points is what makes the definition
diffeomorphism-invariant, by not requiring us to identify points in the two
manifolds; we are comparing instead the two geometries by independently
sampling them, Montecarlo style, which brings a probabilistic aspect into
the definition. Therefore, I begin in section II by briefly reviewing the
definition and some properties of a uniform random distribution of points in a
manifold, with respect to a given volume element. For simplicity, I will
assume that all manifolds $(M,\g)$ have a finite total volume $V_M = \int_M
\dd^\DD {\bf x}\,\rtg$, where $D$ is the dimension of $M$. If the
manifolds have no closed timelike curves, each $n$-point sprinkling is
endowed with a partial order by the causal structure on the manifold, and
defines an element of the set $\C_n$ of all partially ordered sets (posets)
on $n$ points. The idea then is to define $d_n(G,G')$ by comparing the two
probabilities on $\C_n$ corresponding to $G$ and $G'$. The rest of the
section is devoted to constructing a procedure for calculating those
probabilities. Section III contains the definition of the family of
pseudo-distances and a derivation of some of its properties, and section IV
an example in which calculations are carried out in detail. In section V, I
introduce a family of distances, which uses sprinklings of arbitrarily high
numbers of points. The discussion is kept at a general level, independently
of any applications, but the concluding section VI contains additional
remarks on applications of this work as well as open issues.

Finally, a few words concerning notation and terminology. Poset elements are
denoted by $p$, $q$, ...; manifold points by $\bf x$, $\bf y$, ...; by the
{\it past\/} or {\it future\/} of a point $\bf x$ in a Lorentzian manifold,
I will mean its chronological past or future $I^\mp({\bf x})$
(this convention is adopted mainly for the sake of definiteness,
since most of our considerations will depend just on the volume of those sets
or of their intersections and unions, which for well-behaved geometries would
be the same if I had used instead causal pasts/futures or their closures);
and the relationship ${\bf y}\in I^+({\bf x})$, or ${\bf x}\in I^-({\bf
y})$, will be indicated by ${\bf x}<{\bf y}$. Finally, $P$'s will stand for
probabilities, $\tw P$'s for probability densities, $C$'s for posets, $\C$'s
for sets of posets, $V_R$ or $V(R)$ for the volume of the region $R\subseteq
M$, and $R\setminus R' = \{{\bf x}\mid{\bf x}\in R,\; {\bf x}\not\in R'\}$
for the difference between sets.
% \filbreak

\section{II.}{Random Point Distributions and Partially Ordered Sets}

\noindent This section contains the elements that will go into the definition
of the closeness measures. I begin with a summary of the few notions we will
need regarding uniform distributions of points in a manifold, and then
discuss how to obtain probabilities for different resulting partial orders.

Given any manifold $M$ with a volume element, in particular one with a metric
(which at this point could be Riemannian or Lorentzian, possibly even
degenerate---but not everywhere, lest we get $V_M=0$!), such that the total
volume $V_M$ is finite, we can define a random process of sprinkling points
uniformly by stating that, each time a point is chosen in $M$, the probability
density that a particular $\bf x$ be picked is
$$
   \tw P_M({\bf x}|\rtg) = {1\over V_M}\, \sqrt{-g({\bf x})}\;, \eqno(Px)
$$
in any coordinate system; equivalently, the probability that $\bf x$ fall in
any given measurable region $R\subseteq M$ (such as any interval or any finite
union or intersection of such sets \refto{Szab}) is
$$
   P_M({\bf x}\in R)
   = \int_R \tw P_M({\bf x}|\rtg)\,\dd^\DD{\bf x}
   = {V_R\over V_M}\;. \eqno(PxinR)
$$
If the process is repeated $n$ times, we get a uniform, random sprinkling
of points with density $\rho:= n/V_M$, or, if we forget the order they
came in, an unlabelled $n$-point distribution; these are the events we are
interested in, and for which we will calculate probabilities.

One of the probabilities one uses most often in such cases is the one for
exactly $k$ points out of $n$ to fall inside $R$ (without specifying which
ones). This probability follows a binomial distribution,
$$ \eqalignno{
   P(k,R\mid n,M)
   &= {n\choose k} \prod\nolimits_{i=1}^k P({\bf x}_i\in R)
   \,\prod\nolimits_{j=k+1}^n P({\bf x}_j\in M\setminus R) \cr
   &= {n\choose k} \left({V_R\over V_M}\right)^{\!k}
   \left(1-{V_R\over V_M}\right)^{\!n-k}, &(PkinR)}
$$
which, as $V_M$ and $n$ become very large, with $\rho=const$, approaches a
Poisson distribution,
$$
   P(k,R\mid n,M) \approx {\ee^{-\rho V_R}\,(\rho\,V_R)^k\over k!}\;.
   \eqno(Poisson)
$$
This last equation justifies the name {\it Poisson distribution\/} that is
often used for the sets of points used in this paper, and corresponds to the
infinite volume situation. The fact that in that case $P(k,R\mid n,M)$ can be
written in the (exact) form \(Poisson), where only $\rho$ appears and not $n$
or $V_M$, indicates that it may be possible to generalize the definitions and
results of this paper to infinite volumes, although in that case we do not
have the probability density \(Px) available, which is what we would use to
carry out an actual sprinkling, e.g., in a computer simulation.

When we randomly sprinkle $n$ points in $M$, the volume element $\rtg$
determines statistically where they will fall; given their positions, the
causal structure $\hat\g$ determines then the causal relations between them.
From now on, all metrics will have Lorentzian signature and satisfy the past
and future distinguishing condition (see, e.g., Refs \cite{HaEl} and
\cite{Wald}). In particular, this implies the causality condition (no closed
causal curves), which guarantees that a partial order is induced on each
sprinkling, defining an $n$-element poset $C\in\C_n$; the slightly stronger
distinguishing condition implies that ``there are no almost closed causal
curves," in a specific sense which gives some additional benefits, as I will
discuss below. Different geometries $G = \{(M,\g)\}$ and $G' = \{(M',\g')\}$
will then in general give different probabilities $P_n(C|G)$ and $P_n(C|G')$
of obtaining each $C\in\C_n$, which we may compare as a way to determine how
close the geometries themselves are. It is therefore important to have a
general procedure available for calculating, in principle at least, the
probabilities $P_n(C|G)$. 

Let us start by fixing our notation. While $\C_n$ is the set of unlabelled
posets $C$ on $n$ elements, $\ovr\C_n$ will denote the set of labelled
$n$-element posets $\ovr C$, and $\Sigma_n(M)$ the set of $n$-point
sprinklings $\sigma = ({\bf x}_1, \ldots, {\bf x}_n)$ in $M$. (One may argue
that the labelling of the points should not be important; I am considering
sprinklings to be ordered $n$-tuples of points here for convenience.)
As already stated, our random events are $n$-point sprinklings $\sigma$
obtained as a result of a random process with uniform density. The volume
element $\rtg$ on $M$ induces a probability density on $\Sigma_n(M)$; since
the points are independently sprinkled, this can be obtained from products of
single point probability densities \(Px) \refto{Meye},
$$
   \tw P_M({\bf x}_1,\ldots,{\bf x}_n|\rtg)
   = \prod\nolimits_{i=1}^n\tw P_M({\bf x}_i|\rtg)
   = {1\over V_M^n}\prod\nolimits_{i=1}^n\ts{\sqrt{-g({\bf x}_i)}}\;.
   \eqno(Pxi)
$$
If the spacetime $(M,\g)$ has no closed timelike curves, i.e., satisfies the
chronology condition, the relation ${\bf x}_1 < {\bf x}_2$ induced by the
conformal structure $\hat\g$ on $M$ is a partial order, so the sprinkling
$\sigma$ becomes a labelled poset $\ovr C:= \{p_i\mid p_i<p_j\ {\rm iff}\
{\bf x}_i<{\bf x}_j\ {\rm in}\ \sigma\}$, i.e., we get a map $\Phi_{\hat\g}:
\Sigma_n(M) \to \ovr\C_n$ given by $\sigma\mapsto\ovr C$. This map is
many-to-one, and the inverse image of any $\ovr C$ is the set $S =
\Phi_{\hat\g}^{-1}(\ovr C) \subset \Sigma_n(M)$ of all sprinklings with the
same induced labelled partial order. This set has non-zero measure in
$\Sigma_n(M)$; in fact, its probability is
$$
   P_\Sigma(S|\rtg) = \int_S \tw P_M({\bf x}_1,\ldots,{\bf x}_n|\rtg)\,
   \dd^\DD{\bf x}_1\ldots\dd^\DD{\bf x}_n\;, \eqno(PS1)
$$
where $S$ is specified by conditions on the relations between the sprinkled
points giving, for each ${\bf x}_i$, a region $M_i\subset M$ it can fall into
in order to have the right relations with the previously sprinkled ${\bf x}_j$
with $j<i$, according to $\ovr C$. Thus, the probability \(PS1) is of the form
$$
   P_\Sigma(S|\rtg) = {1\over V_M^n}\prod\nolimits_{i=1}^n
   \int_{M_i({\bf x}_1,\ldots,{\bf x}_{i-1};\ovr C)}
   \ts{\sqrt{-g({\bf x}_i)}}\,\dd^\DD{\bf x}_i\;. \eqno(PS2)
$$
This expression gives the probability that the sprinkling give rise to a
labelled poset $\ovr C$; we will see below how to specify the $M_i$ explictly.

What we really want to find is the probability that the sprinkling give rise
to an {\it unlabelled\/} poset $C$. Each $C$ can be labelled in $n!$ ways,
but in general some of these labellings are indistinguishable in terms of the
order relation; more specifically, the number of permutations of elements of
$C$ that give the same $\ovr C$ is the number of automorphisms of $C$,
$|{\rm Aut}(C)|$ (this number is a property of $C$, independent of the
specific $\ovr C$ chosen), and we get that each $C\in\C_n$ can be obtained
from $n!/|{\rm Aut}(C)|$ different labelled $\ovr C$'s, so the probability we
are looking for is
$$
   P_n(C|G):= {n!\over|{\rm Aut}(C)|}\,{1\over V_M^n}
   \prod\nolimits_{i=1}^n \int_{M_i({\bf x}_1,\ldots,{\bf x}_{i-1};\ovr C)}
   \ts{\sqrt{-g({\bf x}_i)}} \,\dd^\DD{\bf x}_i\;, \eqno(PCG)
$$
where $\ovr C$ is an arbitrary labelling of $C$.

Suppose a given labelling $\ovr C = \{p_i\}$ of $C$ has been chosen to carry
out the sprinkling. This means that, in order for the $\{{\bf x}_i\}$ to be a
realization of $\ovr C$, each ${\bf x}_i$ must be in the future of the
${\bf x}_j$'s such that $p_i>p_j$, among the previously sprinkled ones, in
the past of the ones such that $p_i<p_j$, and spacelike related to the
remaining ones. In other words, while ${\bf x}_1$ can be anywhere, $M_1(\ovr
C) = M$, points ${\bf x}_i$ with $i>1$ must fall in the regions
$$
   M_i({\bf x}_1,\ldots,{\bf x}_{i-1};\ovr C)
   = \bigcap_{j<i}\,M_{ij}({\bf x}_j,\ovr C)\;,\qquad
   M_{ij}({\bf x}_j,\ovr C) = \cases{I^+_j &if $p_i>p_j$ \cr
   I^-_j &if $p_i<p_j$ \cr M\setminus I_j &otherwise\ ,} \eqno(Mi)
$$
where for futures and pasts  I use the abbreviations $I^\pm_i:= I^\pm({\bf
x}_i)$ and $I_i:= I^-_i\cup I^+_i$. The most convenient labelling $\ovr C$ to
use in each case may vary. It is often a good choice to pick one compatible
with the partial order on $C$, in the sense that if $p_i<p_j$ then $i<j$,
which can always be done (in fact, it just means ``start labelling from the
bottom and work your way up," and the choice is almost never unique); this
has the advantage that, to reproduce the partial order on $\ovr C$, no ${\bf
x}_i$ needs to be in the past of any of the previously sprinkled ${\bf
x}_j$'s with $j<i$, which eliminates the second case in \(Mi).

This completes the prescription for calculating the probabilities to be used
in the closeness function. In practice, the dependence of each $M_i$ on the
points ${\bf x}_1$, ..., ${\bf x}_{i-1}$ makes the probability very difficult
to calculate analytically, and one would normally use other means such as
computer methods, except for very simple situations like the one in section
IV.

\section{III.}{The Pseudo-Distance}

\noindent In this section, I will take the point of view that geometries $G$
can be identified with the sets of probabilities $\{P_n(C|G)\mid C\in\C_n\}$,
with a degree of approximation that improves as $n$ increases. The task of
defining a pseudo-distance between $G$ and $G'$ is then reduced to that of
defining a distance between their respective sets of probabilities. I will do
so, and then consider some properties of the resulting pseudo-distance.

Various functions can be used as distances between sets of numbers; some
simple ones to handle would be the $\ell^1$-type distance $d_n^{(1)}(G,G')
= \half\,\sum_{C\in\C_n} \left|P_n(C|G)-P_n(C|G')\right|$, the Euclidean
distance, or simply the ``sup" distance, but in view of the interpretation of
the numbers as probabilities, I will use instead the statistical distance
introduced by Wootters \refto{Woot} in the context of rays in Hilbert space,
which is proportional to the number of statistically distinguishable, in an
appropriate sense, intermediate probability sets between the two sets being
compared. Let us then define, for any two geometries,
$$
   d_n(G,G'):= {2\over\pi}\,\arccos\left[\nsum_{C\in\C_n}
   \sqrt{P_n(C|G)}\,\sqrt{P_n(C|G')}\,\right].
   \eqno(dn)
$$
Geometrically, the fact that $\sum_{C\in\C_n}P_n(C|G) = 1$ means that
$\sqrt{P_n(C|G)}$ can be interpreted as the coordinates of a point on the
unit sphere, identifying a direction in probability space $\real^{|\C_n|}$,
and $d_n(G,G')$ is then proportional to the angle defined by the two
corresponding directions; notice that, because all coordinates are
non-negative, that angle is at most $\pi/2$, so with this definition
$d_n(G,G')$ is at most equal to 1.

Clearly, $d_n(G,G')$ is not positive-definite. For each $n$, the number
$|\C_n|$ of posets that can be made out of the $n$ sprinkled points, although
very large, is finite; thus, the value of $d_n(G,G')$ depends on a finite
number of parameters, and cannot capture all of the information contained in
the geometries. This means that $d_n$ cannot be an actual distance function
in the infinite-dimensional space of Lorentzian geometries. One possibility
would be to take the limit $n\to\infty$; this may indeed give a distance, but
it may be a trivial one, as I discuss below, and we shall consider a better
alternative in section V. However, even for finite $n$, two geometries for
which $d_n(G,G') = 0$ are close when probed at scales larger than the mean
point spacing, and this is what we really need in some applications.

Let us consider the other extreme situation, $d_n(G,G') = 1$. For finite $n$,
this can happen only for highly degenerate geometries, since it requires
that the argument of the arccos function in \(dn) be zero, in other words that
there be no $C\in\C_n$ for which both $P_n(C|G)$ and $P_n(C|G')$ are
non-vanishing, i.e., which can be embedded in both geometries. One of the
possible $C$'s is always the totally ordered $n$-element poset (a chain), so
one of the geometries (say, $G$) must assign zero probability to pairs of
timelike related points; in $G$, the light cones of all points must have
degenerated away to lines. Another possible poset is the totally disconnected
one (an antichain), so one of the geometries (necessarily the other one,
$G'$) must assign zero probability to pairs of spacelike related points; $G'$
has the wide open light cones of the infinite speed of light limit, or is a
one-dimensional timelike line. No poset $C\in\C_n$ other than those two can be
embedded in either geometry. We conclude that the inequality $d_n(G,G') \le
1$ cannot be saturated other than as a degenerate limit of sequences of
geometries of the type just described.

In the limit $n \to \infty$, however, the situation may change. We know from
continuum results that the topology, differentiable, and conformal structures
of a past and future distinguishing Lorentzian geometry can be recovered just
from the knowledge of the causal relations between {\it all\/} pairs of points
\refto{HKMC, Mala} and that, if one considers instead pairs of points in a
sequence of uniform sprinklings of increasing density, the same is true in
the limit $n \to \infty$, with the added bonus that the volume element can be
recovered as well, up to a global factor \refto{MeSo, BoMe}; thus, in that
limit, sequences of posets $\{C_n\}$, where each $C_n$ has $n$ elements and
is a subposet of the next one, $C_n \subset C_{n+1}$, can be embedded at most
in a single geometry $G$. This means that
$$
   \forall\,G\ne G',\quad
   \lim_{n\to\infty}P_n(C_n|G)\,P_n(C_n|G') = 0
   \quad \forall\,\{C_n\}\;. \eqno(sequence)
$$
In fact, it is also true that each individual probability $P_n(C_n|G)$ or
$P_n(C_n|G')$ tends to zero as $n \to \infty$. But the number of terms in
the summation in \(dn) grows very fast with $n$ (faster than exponentially
\refto{KlRo}), and the limiting value of $d_n(G,G')$ for $G \ne G'$ depends on
the rate of approach to zero of these probabilities. It is possible that
$d_\infty(G,G') = 1$ for all $G \ne G'$ (in terms of the discussion above,
many posets may be embeddable in both $G$ and $G'$, but the limit is 1
because all products of probabilities in \(sequence) go to zero fast enough);
in this case $d_\infty$ would be a distance, but a trivial, not very useful
one.

We will see what $d_\infty(G,G')$ can be replaced by later in the paper, and
return now to examing properties of $d_n$ with finite $n$. In addition to its
much greater ease of computation, the function $d_n(G,G')$ is also made
interesting by the following reasonable conjectures: (i) In a sense, for large
enough $n$, it is ``almost a distance," or ``positive-definite up to
differences on small scales;" (ii) For any subset of geometries labelled by a
finite number $N$ of parameters (analogous to the ``minisuperspaces" used for
spatial geometries), there is a finite $n$ such that $d_n$ is a true distance
function on this set, and (iii) For any two arbitrary (distinguishing,
finite-volume) different geometries $G$ and $G'$ there is a finite $n$ such
that $d_n(G,G')>0$, with $d_n(G,G') \to 1$ as $n\to\infty$.

To start with, I will prove the intuitively obvious, and nice property of the
closeness measure that it is a monotonically increasing function of $n$:
$$
   \forall\,G,\,G'\qquad d_n(G,G')\le d_{n+1}(G,G')\;. \eqno(lemma)
$$
To prove this inequality, consider the process of sprinkling $n+1$ points in a
geometry $G$ as an $n$-point sprinkling, followed by the choice of one more
point. Then, the probability of the first $n$ points yielding any given
$C\in\C_n$ is a sum over probabilities for different $C'\in\C_{n+1}$ obtained
when the extra point is added,
$$
   P_n(C|G) = \nsum_{C'\in\C_{n+1}}f_{C,C'}\,P_{n+1}(C'|G)\;,
   \qquad f_{C,C'}:= {1\over n+1}\,{C'\choose C}\;, \eqno(choose)
$$
where $f_{C,C'}$ is the fraction of $n$-element subsets of $C'$ that are
isomorphic to $C$, which can be expressed in terms of the number ${C'\choose
C}$ of ways of picking an $n$-element subset of $C'$ that is isomorphic to $C$
(this number may be called ``$C'$ choose $C$," and I will use the convention
that it vanishes if $C$ is not a subposet of $C'$); notice that it is clear
from the definition that $\sum_{C\in\C_n}f_{C,C'} = 1$, for any $C'$. Then,
we can write
$$
   d_n(G,G') = {2\over\pi}\,\arccos\left[\nsum_C
   \sqrt{\nsum_{C'} f_{C,C'}\,P_{n+1}(C'|G)}\,
   \sqrt{\nsum_{C''} f_{C,C''}\,P_{n+1}(C''|G')}\,\right], \eqno(proof1)
$$
where it is understood that $C\in\C_n$ and $C',C''\in\C_{n+1}$. For each $C$,
the corresponding term in the summation in \(proof1) is of the form
$\sqrt{(\sum_ia_i)(\sum_jb_j)}$ where all $a_i$ and $b_j$ are non-negative,
for which the general inequality
$$
   \sqrt{\Big(\ts\sum_ia_i\Big)\Big(\ts\sum_jb_j\Big)}
   \ge \nsum_i\sqrt{a_ib_i} \eqno(proof2)
$$
holds. To prove this inequality, we can square the two sides, which gives
$\sum_i\sum_ja_ib_j$ and $\sum_i\sum_j\sqrt{a_ib_i}\sqrt{a_jb_j}$,
respectively; the terms with $i=j$ are equal; separate the other ones in
pairs, $a_ib_j + a_jb_i$ and $2\sqrt{a_ib_i}\sqrt{a_jb_j}$, respectively, and
square them; since we always have $a_i^2b_j^2 + 2\,a_ia_jb_ib_j + a_j^2b_i^2
\ge 4\,a_ia_jb_ib_j$, \(proof2) follows. Applying this to \(proof1) gives
$$ \eqalignno{
   d_n(G,G') &\le {2\over\pi}\,\arccos\left[\nsum_C\nsum_{C'}
   \sqrt{(f_{C,C'})^2 P_{n+1}(C'|G)\,P_{n+1}(C'|G')}\,\right] \cr
   &= {2\over\pi}\,\arccos\left[ \nsum_{C'}\Big(\nsum_Cf_{C,C'}\Big)
   \sqrt{P_{n+1}(C'|G)\,P_{n+1}(C'|G')}\,\right] \cr
   &= d_{n+1}(G,G')\;. &(proof3) }
$$
As a consequence of the proof, we also see that
$$
   d_n(C,C') = d_{n+1}(C,C')\qquad{\rm iff}\qquad d_{n+1}(C,C') = 0\;,
   \eqno(zero)
$$
since the inequality in \(proof3) can only be saturated if \(proof2) is, and
this will happen only if for all $i$ and $j$, $a_ib_j = a_jb_i$, which in
terms of our probabilities reduces to $P_{n+1}(C'|G) = P_{n+1}(C'|G')$.
As a byproduct, we also obtain the equality \(choose), which may be useful
for calculating $P_n(C|G)$, or one of the $P_{n+1}(C'|G)$'s if the others
are known.

\section{IV.}{A Simple Example}

\noindent As an illustration of the definition of $d_n(G,G')$ and the
procedure for calculating $P_n(C|G)$ introduced in section II, we consider a
very simple example, which already involves two 1-parameter families of
geometries with different underlying manifolds: a finite-size rectangular
portion of 2-dimensional Minkowski space, with line element $\dd s^2 = -\dd
t^2 + \dd x^2$ and topology $M \simeq \real^2$, $G_\gamma = \{(M,\eta)\}$, and
a similar one obtained after a spatial identification, with the same line
element and topology $M' \simeq \real \times {\rm S}^1$, $G_\delta =
\{(M',\eta)\}$. I will first introduce each geometry and calculate the
simplest probabilites, $P_2(C|G_\gamma)$ and $P_2(C|G'_\delta)$, then use
these to find $d_2(G_\gamma,G'_\delta)$; the results will give us an
indication of features and limitations of $d_2$, and we will then see how to
overcome these limitations by calculating the $P_3$'s and using
$d_3(G_\gamma,G'_\delta)$.

The geometry $G_\gamma$ is the rectangle $M:=\{{\bf x}\mid 0\le x\le a,\,
0\le t\le b\}$ in two-dimensional Minkowski space. Since the probabilities
we are looking for are invariant under a global rescaling, they cannot depend
on the volume $V_M=ab$, but only on the aspect ratio $\gamma:= b/a$. For $n =
2$, $\C_2$ has two elements, the connected two-element poset $\conn$ and the
disconnected one $\disc$; we must calculate the integrals in \(PCG) for
these two posets.

To get the connected poset $\conn$ in a two-point sprinkling with the
``bottom-up" labelling, we need ${\bf x}_2$ to fall in the future of
${\bf x}_1$, or $M_2({\bf x}_1,\ovr C) = M_{21}({\bf x}_1,\ovr C) = I^+_1$,
and \(PCG) becomes
$$
   P_2(\conn\,|G_\gamma)
   = {2!\over1\cdot V_M^2}\int_M\!\dd^2{\bf x}_1\int_{I^+_1}\!\dd^2{\bf x}_2
   = {2\over(ab)^2}\int_0^a\!\dd x\int_0^b\!\dd t \;V(I^+(x,t))\;,\eqno(ex.1)
$$
where ${\bf x}_1 = (x,t)$, and the volume $V(I^+(x,t))$ is $(b-t)^2$, with
correction terms that are needed for some values of $(x,t)$ (see Fig.\ 1),
$$
   \eqalignno{
   V(I^+_1) = (b-t)^2 &- {(b-t-x)^2\over2}
   \,\theta(x<b\ {\rm and}\ t < b-x) \cr
   &- {(b-t-a+x)^2\over2}
   \,\theta(x>a-b\ {\rm and}\ t<b-a+x)\;. &(VIplus)}
$$
(The step function $\theta$ equals 1 if its argument is true, 0 otherwise.)
If we assume that $\gamma \ge 1$, so that $x<b$ and $x>a-b$ are always
satisfied, \(ex.1) gives
$$
   P_2(\conn\,|G_\gamma)
   = {1-4\,\gamma+6\,\gamma^2\over6\,\gamma^2}\;, \eqno(Pconn)
$$
To get the disconnected poset $\disc$ we need the points to be causally
unrelated, $M_2({\bf x}_1,\ovr C) = M\setminus I_1$, so \(PCG) becomes
$$
   P_2(\disc\,|G_\gamma)
   = {2!\over2\cdot V_M^2}\int_M\!\dd^2{\bf x}_1
   \int_{M\setminus I_1}\!\dd^2{\bf x}_2
   = 1 - P_2(\conn\,|G) = {4\,\gamma-1\over6\,\gamma^2}\;, \eqno(Pdisc)
$$
where instead of doing another integral I have used $P_2(\conn\,|G_\gamma)
+ P_2(\disc\,|G_\gamma) = 1$.

For the case $b < a$, we can now either integrate \(VIplus) again, or use
simple symmetry considerations. If we flip the rectangle by exchanging $a
\leftrightarrow b$, the manifold transforms according to $G_\gamma
\leftrightarrow G_{1/\gamma}$; if we take the two sprinkled points
${\bf x}_1$ and ${\bf x}_2$ along, the posets are also turned into each
other, $\conn\leftrightarrow\disc$. Thus,
$$
   \eqalign{
   P_2(\conn\,|G_\gamma) &= P_2(\disc\,|G_{1/\gamma})
   = {4\gamma-\gamma^2\over6} \cr
   P_2(\disc\,|G_\gamma) &= P_2(\conn\,|G_{1/\gamma})
   = {\gamma^2-4\gamma+6\over6}\;. \cr} \eqno(trick)
$$

The geometry $G'_\gamma$ is the cylinder $M'$ one obtains applying the spatial
identification $(t,0) \sim (t,a)$ to the rectangle in $G_\gamma$, with the
same line element; again, the probabilities only depend on the aspect ratio
$\gamma:= b/a$. Similar calculations to the ones leading to \(Pconn) and
\(Pdisc), but now integrating
$$
   V(I_1'{}^+)
   = (b-t)^2 - (b-t-\half\,a)^2\,\theta(t<b-\half\,a) \eqno(VIplusprime)
$$
(see Fig.\ 2) over $M'$, give
$$
  P_2(\conn\,|G'_\gamma)
  = {1-6\,\gamma+12\,\gamma^2\over12\,\gamma^2}\;, \qquad
  P_2(\disc\,|G'_\gamma)
  = {6\,\gamma-1\over12\,\gamma^2}\;, \eqno(ex.3)
$$
for $\gamma\ge\half$, and
$$
  P_2(\conn\,|G'_\gamma) = \ts{2\over3}\,\gamma\;, \qquad
  P_2(\disc\,|G'_\gamma) = 1 - \ts{2\over3}\,\gamma\;, \eqno(ex.4)
$$
for $\gamma\le\half$, when $V(I_1'{}^+)$ is just $(b-t)^2$; we cannot use a
trick like the one in \(trick) here, but this probability is the easiest one
to calculate anyway.

If we now use the definition \(dn) to calculate
$$
   d_2(G_\gamma,G'_\gamma) = {2\over\pi}\,\arccos\left[
   \sqrt{P_2(\conn\,|G_\gamma)\,P_2(\conn\,|G'_\gamma)}
   + \sqrt{P_2(\disc\,|G_\gamma)\,P_2(\disc\,|G'_\gamma)}
   \,\right], \eqno(ex.5)
$$
where the two geometries are characterized by the same parameter value
$\gamma$, we get the function plotted in Fig.\ 3, which goes to zero as the
aspect ratio $\gamma \to 0$ or $\infty$, and the difference between the two
manifolds becomes immaterial because all pairs $({\bf x}_1,{\bf x}_2)$ are
spacelike or timelike related, respectively, in both geometries; but
$d_2(G_\gamma,G'_\gamma)$ is not zero for any non-degenerate cases. However,
if we use $d_2$ for geometries with different parameter values, we find that,
for example,
$$
   {\rm if}\ \gamma,\;\delta \le \half\;,\qquad
   P_2(\conn\,|G_\gamma) = P_2(\conn\,|G'_\delta)
   \quad {\rm when}\quad \delta = \gamma-\ts{1\over4}\,\gamma^2\;,
   \eqno(ex.6)
$$
i.e., for each $\gamma$ there is a $\delta$ such that
$d_2(G_\gamma,G'_\delta)$ vanishes; with two sprinkled points, the single
available parameter $P_2(\conn|G)$ cannot distinguish all geometries in
this example. We will now see that this can be done using three points.

What we need to show is that, for all values of the parameters, among the
elements of $\C_3 = \{\tre, \due, \vee, \eev, \lin\}$, there is at least one
which is embeddable in $G$ and $G'$ with different probabilities; because of
property \(lemma), we actually only need to do this for the parameter values
for which $d_2(G_\gamma, G'_\delta) = 0$. Let us consider the three-element
poset for which the probabilities are easiest to calculate, the linear order
$\lin$ (linear orders are always the easiest ones, because $M_{ij}({\bf
x}_j,\ovr C)$ reduces just to $I_j^+$ and $M_i({\bf x}_1,...,{\bf
x}_{i-1};\ovr C)$ to $I_{i-1}^+$ in \(Mi), if $\ovr C$ is the ``bottom-up"
labelling). The calculations again proceed along the lines of those for
\(Pconn) but are somewhat longer, since we now have to evaluate
$$
   P_3(\lin\,|G) = {3!\over1\cdot V_M^3} \int_M\!\dd^2{\bf x}_1
   \int_{I^+_1}\!\dd^2{\bf x}_2 
   \int_{I^+_2}\!\dd^2{\bf x}_3\;. \eqno(ex.7)
$$
I will restrict myself to the case $\gamma \le \half$, where by explicitly
integrating \(ex.7) one gets
$$
   P_3(\lin\,|G_\gamma)
   = \ts{1\over5}\,\gamma^2 - \ts{1\over12}\,\gamma^3\;,
   \qquad P_3(\lin\,|G'_\gamma) = \ts{1\over5}\,\gamma^2\;.
   \eqno(ex.8)
$$
When $\delta = \gamma-{1\over4}\gamma^2$, a simple calculation gives
$$
   P_3(\lin\,|G_\gamma) - P_3(\lin\,|G'_\delta)
   = \ts{1\over20}\,\gamma^3\,(\ts{1\over3}-\ts{1\over4}\,\gamma)\;,
   \eqno(ex.9)
$$
which does not vanish for $\gamma \le \half$, so $d_3(G_\gamma,G'_\delta) \ne
0$ as expected. The two families of geometries are different enough that the
induced order on a random three-element subset will pick out the difference.
The other three-point probabilities $P_3(C|G)$ can be found without too much
additional effort. I will now show how to do this, since the information
those probabilities capture about the geometries is interesting in its own
right, and because this will illustrate the use of some of the general
relationships introduced above, as well as one new relationship and possible
symmetries.

Suppose we want to calculate the values of the probabilities
$P_3(\tre\,|G)$, $P_3(\due\,|G)$, $P_3(\vee\,|G)$, $P_3(\eev\,|G)$, and
$P_3(\lin\,|G)$ for some geometry $G$, for which we already know the values
of $P_2(\disc\,|G)$ and $P_2(\conn\,|G)$; this includes having calculated
$V(I^+_i)$ for every ${\bf x}_i\in M$, as in \(VIplus), or some other similar
integral. It would be to our advantage to use as many relationships as
possible among the $P_3(C|G)$'s. One is always given by the identity
$$
   \nsum_{C\in\C_3}P_3(C|G) = 1\;, \eqno(rel1)
$$
and two more are always given by
$$ \eqalignno{   
   &\ts{1\over3}\,P_3(\due\,|G) + \ts{2\over3}\,P_3(\vee\,|G) +
   \ts{2\over3}\,P_3(\eev\,|G) + P_3(\lin\,|G)
   = P_2(\conn\,|G) \;, &(rel2) \cr
   & P_3(\tre\,|G) + \ts{2\over3}\,P_3(\due\,|G) + \ts{1\over3}
   \,P_3(\vee\,|G) + \ts{1\over3}\,P_3(\eev\,|G)
   = P_2(\disc\,|G) \;, &(rel3) }
$$
arising from \(choose). Notice however that only two of the three
relationships \(rel1)--\(rel3) are independent, since \(choose) already
implies $\sum_{C'\in\C_{n+1}} P_{n+1}(C'|G) = 1$ if one uses valid
$P_n$'s and $f_{C,C'}$'s, satisfying $\sum_{C\in\C_n} P_n(C|G) = 1$
and $\sum_{C\in\C_n} f_{C,C'} = 1$. An additional relationship, not as simple
as \(rel2)--\(rel3) but still useful, can be found by considering
probabilities for subsets of $\C_3$ rather than just single $P_3(C|G)$'s.
Since the outcomes $\vee$ and $\lin$ are mutually exclusive,
$$
   P_3(\vee\ {\rm or}\ \lin\,|G)
   = P_3(\vee\,|G) + P_3(\lin\,|G)\;, \eqno(veep.1)
$$
which means that
$$
   \eqalignno{
   P_3(\vee\,|G) &= P_3(\vee\ {\rm or}\ \lin\,|G) - P_3(\lin\,|G) \cr
   &= {3!\over2\cdot V_M^3} \int_M\!\dd^2{\bf x}_1
   \int_{I^+_1}\!\dd^2{\bf x}_2 \int_{I^+_1}\!\dd^2{\bf x}_3 - P_3(\lin\,|G)
   \cr &= {3\over V_M^3} \int_M\!\dd^2{\bf x}\,
   \bigl(V(I^+(x,t))\bigr)^{\!2} - P_3(\lin\,|G)\;, &(veep.2)}
$$
where I have used the fact that, in a ``bottom-up" labelling, the condition
for obtaining $\vee$ or $\lin$ is simply that both ${\bf x}_2$ and ${\bf x}_3$
be in the future of ${\bf x}_1$. Finally, if $G$ was time reversal invariant
(the $G_\gamma$ and $G'_\delta$ in our example both are), we would get an
additional relationship,
$$
   P_3(\eev\,|G) = P_3(\vee\,|G)\;. \eqno(rel4)
$$
We have found four relationships among the $P_3(C|G)$'s; if they hold,
only one probability needs to be calculated by direct application of \(PCG).
Also, in specific cases, it may be possible to use other symmetries of $G$ to
derive relationships of other types; for example, the one I used in the trick
of \(trick) involving different parameter values.

In our example, we start by calculating the integral in \(veep.2) using
\(VIplus); if $\gamma < \half$,
$$
   \eqalignno{
   \int_M\dd^2{\bf x}\,\bigl(V(I^+(x,t))\bigr)^{\!2}
   &= \int_0^a\dd x \int_0^b\dd t\,(b-t)^4 \cr
   &\phantom= + \int_0^b\dd x\int_0^{b-x}\dd t
   \left[{(b-t-x)^4\over4}-(b-t)^2\,(b-t-x)^2\right] \cr
   &\phantom= + \int_{a-b}^a\dd x\int_0^{b-a+x}\dd t
   \left[{(b-t-a+x)^4\over4}-(b-t)^2\,(b-t-a+x)^2\right] \cr
   &=\vphantom{\int} \ts{1\over5}\,ab^5 - \ts{17\over180}\,b^6\;, &(int)}
$$
where one of the cross terms in the square of \(VIplus) does not contribute
because in this case the intervals $x<b$ and $x>a-b$ don't overlap; then,
using $P_3(\lin\,|G)$ from \(ex.8), we find $P_3(\vee\,|G)$, and from \(rel4)
we find $P_3(\eev\,|G)$; with these, \(rel2) gives $P_3(\due\,|G)$, and
\(rel3) gives $P_3(\tre\,|G)$. Analogous calculations can be done for $G'$,
starting with
$$
   \int_{M'}\dd^2{\bf x}_1\bigl(V(I^+_i)\bigr)^{\!2}
   = \int_0^a\dd x\int_0^b\dd t\,(b-t)^4
   = \ts{1\over5}\,ab^5\;.\eqno(intprime)
$$
To conclude, the full sets of probabilities we have calculated for
$\gamma<\half$, are
\medskip
\settabs\+\hskip40pt&\hskip200pt&\hskip180pt \cr
\+ &$P_2(\conn\,|G_\gamma) = {2\over3}\,\gamma-{1\over6}\,\gamma^2$
   &$P_2(\conn\,|G'_\gamma) = {2\over3}\,\gamma$ \cr
\smallskip
\+ &$P_2(\disc\,|G_\gamma) = 1-{2\over3}\,\gamma+{1\over6}\,\gamma^2$
   &$P_2(\disc\,|G'_\gamma) = 1-{2\over3}\,\gamma$ \cr
\medskip
\noindent for two-point sprinklings, and
\medskip
\+ &$P_3(\lin\,|G_\gamma) = {1\over5}\,\gamma^2-{1\over12}\,\gamma^3$
   &$P_3(\lin\,|G'_\gamma) = {1\over5}\,\gamma^2$ \cr
\smallskip
\+ &$P_3(\vee\,|G_\gamma) = {2\over5}\,\gamma^2-{1\over5}\,\gamma^3$
   &$P_3(\vee\,|G'_\gamma) = {2\over5}\,\gamma^2$ \cr
\smallskip
\+ &$P_3(\eev\,|G_\gamma) = {2\over5}\,\gamma^2-{1\over5}\,\gamma^3$
   &$P_3(\eev\,|G'_\gamma) = {2\over5}\,\gamma^2$ \cr
\smallskip
\+ &$P_3(\due\,|G_\gamma)
   = 2\,\gamma-{27\over10}\,\gamma^2+{21\over20}\,\gamma^3$
   &$P_3(\due\,|G'_\gamma) = 2\,\gamma-{11\over5}\,\gamma^2$ \cr
\smallskip
\+ &$P_3(\tre\,|G_\gamma)
   = 1-2\,\gamma+{17\over10}\,\gamma^2-{17\over30}\,\gamma^3$
   &$P_3(\tre\,|G'_\gamma) = 1-2\,\gamma+{6\over5}\,\gamma^2$ \cr
\medskip
\noindent for three-point sprinklings. The corresponding ones for
$\gamma>\half$ can be similarly calculated with the formalism
described above.

\section{V.}{The Distance Function}

\noindent We can now extend the definition of the closeness function to a
distance. Only the definition and a few comments will be given here; a more
extensive study of its properties is left for future work. From the previous
discussion, it should be clear that in this case we need to let the number of
points sprinkled in each manifold go to infinity, so that we probe its
structure at arbitrarily small scales. Also, it is not sufficient to let
$n\to\infty$ in $d_n(G,G')$, both because the resulting distance may be
trivial, and because, like all $d_n$'s, it would not distinguish between
different values of the total volumes $V_M$ and $V_{M'}$. Each random event
considered in previous sections was the choice of $n$ points in each
manifold; since $n$ was the same for both manifolds, the outcomes gave us no
information on their total volumes. We can overcome this limitation in the
distance by letting each random event consist in the choice of both $n$ and
the location of the points, and this will give me an opportunity to mention
one feature of the closeness functions that had not explicitly come up until
now.

Consider two geometries $G = \{(M,g)\}$ and $G' = \{(M',g')\}$, as before.
We will draw the number of points $n$ to be sprinkled in each manifold from
Poisson distributions, whose means will be proportional to the respective
volumes, and will thus in general be different, the relationship between the
two means being given by the point densities they correspond to. However,
the two manifolds may be of different dimensions, $D$ and $D'$ respectively,
in which case it would be meaningless to require the two volume densities of
points to be equal; what we can require is equality of the ``mean point
spacings", i.e., that the volume densities $\rho$ and $\rho'$ satisfy
$(\rho)^{1/\DD} = (\rho')^{1/\DD'} = \ell^{-1}$. (For example, in quantum
gravity applications, we can think of $\ell$ as being the Planck length $\lp$,
and the issue of different dimensionalities is relevant for higher-dimensional
theories such as the Kaluza-Klein ones ---for a recent review, see Ref\
\cite{OvWe}--- where we might want to compare a macroscopic four-dimensional
manifold $^4\!M$ to a $D$-dimensional fundamental one which is, at least
locally, considered to be a product of the type $^\DD\!M \simeq {}^4\!M
\times {}^{\DD-4}\!M$, with $^{\DD-4}\!M$ of volume $\lp^{\DD-4}$.)

To define the distance function, choose a positive mean point spacing $\ell$
around which most of the contribution to the distance will come from; sprinkle
points in $G$ and $G'$ by first choosing, each time, the number of points
according to Poisson distributions
$$
   P_\mu(n) = {\ee^{-\mu}\mu^n\over n!}\;, \qquad
   P_{\mu'}(n) = {\ee^{-\mu'}\mu'{}^n\over n!}\;,
   \eqno(Pn)
$$
respectively, where $\mu:= V_M/\ell^\DD$ and $\mu':= V_{M'}/\ell^{\DD'}$, and
distribute in $(M,g)$ and $(M',g')$ the chosen numbers of points uniformly
at random; the probabilities of obtaining any given poset $C\in\C_n$ as a
result are now, respectively,
$$
   P_\ell(n,C|G) = P_\mu(n)\,P_n(C\mid G)\;,\qquad
   P_\ell(n,C|G') = P_{\mu'}(n)\,P_n(C\mid G')\;; \eqno(Prho)
$$
finally, compare these probabilities by extending \(dn) to
$$
   \eqalignno{
   d_\ell(G,G'):&= {2\over\pi}\,\arccos\Biggl[\sum_{n=0}^\infty\,
   \sum_{C\in\C_n} \sqrt{P_\ell(n,C|G)}\,\sqrt{P_\ell(n,C|G')}\,\Biggr]\cr
   &= {2\over\pi}\,\arccos\Biggl[\sum_{n=0}^\infty
   \biggl(\sqrt{P_\mu(n)\,P_{\mu'}(n)}
   \,\sum_{C\in\C_n} \sqrt{P_n(C|G)\,P_n(C|G')}\biggr)
   \,\Biggr]. &(dell)}
$$
Here, I am assuming we have defined the probabilities $P_n(C|G)$ for the
sets of one-element posets, $\C_1 = \{\uno\}$, and zero-element posets,
$\C_0 = \{\emptyset\}$; if we set $P_1(\uno|G) = 1$, consistently with the
general definition, and adopt the convention that $P_0(\emptyset |G) = 1$,
the argument of the arccos function can be written as
$$
   \sqrt{\vphantom{P_\mu}\ee^{-\mu}\,\ee^{-\mu'}}
   + \sqrt{\vphantom{P_\mu}\mu\mu'\,\ee^{-\mu}\,\ee^{-\mu'}}
   + \sum_{n=2}^\infty \biggl(\sqrt{P_\mu(n)\,P_{\mu'}(n)}
   \,\sum_{C\in\C_n} \sqrt{P_n(C|G)\,P_n(C|G')}\biggr)\;. \eqno(arg)
$$

The expressions \(dell) and \(arg) are clearly well-defined; the rapid
decrease of $P_\mu(n)$ and $P_{\mu'}(n)$ for large $n$ makes them finite, and
the fact that they are probabilities implies that \(dell) actually gives a
number between 0 and 1, for the same reason as the pseudo-distance \(dn) did.
To examine these two extreme situations, consider \(arg). This expression
vanishes only in the large $\mu$ or $\mu'$ limit ($\ell\to0$, or at least
one of the volumes $\to\infty$), so that the first two terms vanish, and if
the contribution from all $n\ge2$ vanishes; this may imply that the conformal
structures have the degeneracies described earlier in section III for
$d_n(G,G')$, at least if the manifolds have equal volumes, but in any case
we can already see that $d_\ell(G,G) = 1$ only in situations obtained as 
limits of ones of the type under consideration.

The more interesting situation is when $d_\ell(G,G') = 0$. We can see from
\(dell) that this implies $P_n(C|G) = P_n(C|G')$ for all $n$ and $C$, since
for all $n$ the summation over $C\in\C_n$ must equal 1, and $P_\mu(n) =
P_{\mu'}(n)$ for all $n$, since the sum over $n$ also must give 1. The latter
conditions obviously mean that $V_M = V_{M'}$, and the former set implies $G
= G'$; the sketch of a proof goes as follows. When we sprinkle an increasing
number of points in a manifold, we build a sequence of posets
$\{C_n\}_{n\in\NN}$, with $C_n\subset C_{n+1}$ for all $n$. In the limit
$n\to\infty$, we obtain the direct limit of this sequence, $C_\omega:=
\bigcup_{n=1}^\infty C_n$, and the equality of all $P_n(C|G) = P_n(C|G')$
turns into the equality of appropriately defined probabilities $P_\infty(C|G)$
and $P_\infty(C|G')$ on $\C_\infty$. Now, if we apply a completion procedure
to $C_\omega$, analogous to the Dedekind cut construction of real numbers
from rational ones, we get the points of the original manifold together with
their causal relations and the conformal factor, i.e., we get back the
geometry $G$ \refto{BoMe,MeSo}. But if the two probabilities $P_\infty(C|G)$
and $P_\infty(C|G')$ are equal, and the completion of infinite posets drawn
from them gives respectively $G$ and $G'$, with probability one, it must be
that $G = G'$, and thus $d_\ell$ is positive-definite.

\section{VI.}{Concluding Remarks}

\noindent To measure the closeness of Lorentzian geometries, I have
introduced a family of pseudo-distances $d_n(G,G')$ and a family of distances
$d_\ell(G,G')$ on the space of all past and future distinguishing Lorentzian
geometries of finite volume. The main idea was to sprinkle points uniformly
at random in $G$ and $G'$, and use the resulting probabilities $P_n(C|G)$ as
the basic ingredients for the functions; in this paper, those probabilities
were combined using one specific distance between probability measures, but
others are known and may be more suitable in some applications. The closeness
functions presented here, together with other possible such functions based
on the same probabilities, are the only non-trivial diffeomorphism-invariant
ones on this space that I am aware of.

A number of interesting questions arise about the statistical approach to
Lorentzian geometry discussed here. Even before their use in the closeness
functions, the probabilities $P_n(C|G)$ are interesting in themselves, as a
complete set of invariants (together with the volume) of finite volume,
distinguishing Lorentzian geometries. It would be worth while to study the
type of information about the manifold that those different invariants
contain; for example, how they encode dimensionality, how they are affected
by conformal transformations as opposed to changes in the conformal
structure, or how one can tell ``localized" changes from ``global" changes in
a manifold from their effect on the $P_n(C|G)$. Possible starting points in
answering these questions may consist in examining examples along the lines
of the one in section IV but in which different parameters are varied, e.g.,
comparing a two-dimensional and a three-dimensional manifold, or modifying
one by a conformal transformation; and studying analytically the effect of
small variations $\g \mapsto \g+\delta\g$.

The answer to questions of the above type may then allow us to word in a
more precise way statements like ``geometries for which $d_n(G,G')$ is small
are close down to the scale $V_M/n$;" understand how the topology induced
by the $d_n$'s and $d_\ell$'s on the space of Lorentzian geometries relates
to previously studied ones \refto{BeEh}; and place bounds on the value of
$d_n(G,G')$ when the actual value cannot be calculated, for example through 
bounds on the probabilities $P_n$ due to the non-embeddability of some $C$'s
in a geometry, such as $C$'s that require higher dimensions. The infinite
density limit and the properties of $d_\ell$ need to be understood better
than what is sketched in section V, and the limit in which the ``regulator"
$\ell$ is taken to zero is a potentially useful one. It would also be useful
to extend the present work to a definition of closeness that applies to
infinite volume manifolds, as mentioned in section II; in that case, one may
need to introduce a quasi-local element in the definition, and use finite
size subsets of sprinklings of density $\rho$.

On the physical side, this work may be related to definitions of approximate
solutions of Einstein's equation \refto{Bel, GaVe, DeBr}, and spacetimes with
approximate symmetries \refto{Zala}, which have been considered for various
reasons, including their relevance to the issue of gravitational entropy and
the smoothing problem in cosmology \refto{Ser2}. These problems, in addition
to the motivation coming from quantum gravity, make it an interesting issue to
study properties of $P_n(C|G)$, by analytical methods or numerical
simulations.

%\vfill\eject
\section{Acknowledgements}{}

\noindent I would like to thank Sebastiano Sonego for many useful and
stimulating discussions, David Meyer and Rafael Sorkin for helpful remarks,
and an anonymous referee for suggestions that led to significant improvements
in the content of this paper. Part of this work was done at the Universit\'e
Libre de Bruxelles, where it was supported by the Directorate-General for
Science, Research and Development of the Commission of the European
Communities under contract no.\ CI1*-0540-M(TT); and at the Department of
Theoretical Physics of the Universit\'e de Gen\`eve, whose hospitality is
gratefully acknowledged.

\section{References}{}
\baselineskip=12pt
\parskip=0pt\parindent=25pt
\frenchspacing
\leftskip=30pt\parindent=-10pt
\def\new{\hfill\break\indent}\def\dash{---\hskip-1pt---}

\refis{Asht}A. Ashtekar, ``Quantum mechanics of geometry," gr-qc/9901023.

\refis{Baez}J.C. Baez, ``An introduction to spin foam models of quantum
gravity and BF theory," gr-qc/9905087.

\refis{BeEh}J.K. Beem and P.E. Ehrlich, {\sl Global Lorentzian Geometry\/}
(Dekker 1981).

\refis{Bel}L. Bel, ``The quality factor of approximate solutions of
Einstein's equations," {\sl Gen. Rel. Grav.} {\bf19}, 1127-1130 (1987).

\refis{BLMS}L. Bombelli, J. Lee, D.A. Meyer and R.D. Sorkin, ``Spacetime as a
causal set," {\sl Phys. Rev. Lett.} {\bf59}, 521-524 (1987); see also C.
Moore, ``Comment to `Spacetime as a causal set'{}'' (and reply), {\sl Phys.
Rev. Lett.} {\bf60}, 655-656 (1988).

\refis{Bomb}L. Bombelli, {\sl Spacetime as a Causal Set}, PhD thesis, Syracuse
University 1987.

\refis{BoMe}L. Bombelli and D.A. Meyer, ``The origin of Lorentzian geometry,"
{\sl Phys. Lett.} {\bf141A}, 226-228 (1989).

\refis{BoSo}L. Bombelli and R.D. Sorkin, unpublished work.

\refis{BrGr}G. Brightwell and R. Gregory, ``Structure of random discrete
spacetime," {\sl Phys. Rev. Lett.} {\bf66}, 260-263 (1991).

\refis{BSal}L. Bombelli, ``Causal sets and the closeness of Lorentzian
manifolds," in {\sl Relativity in General\/} (Spanish Relativity Meeting
1993), J. D\'{\i}az Alonso and M. Lorente P\'aramo, eds. (Editions
Fronti\`ere 1994).

\refis{CrSm}L. Crane and L. Smolin, ``Spacetime foam as a universal regulator"
{\sl Gen. Rel. Grav.\/} {\bf17}, 1209 (1985).

\refis{DeBr}S. Detweiler and L.H. Brown, ``The post Minkowskii expansion of
general relativity," {\sl Phys. Rev.} D {\bf56}, 826-841 (1997), and
gr-qc/9609010.

\refis{DePi}R. De Pietri, ``Canonical loop quantum gravity and spin foam
models," gr-qc/9903076.

\refis{GaVe}J. Garriga and E. Verdaguer, ``Testing the quality factor with
some approximate solutions to Einstein's eqs," {\sl Gen. Rel. Grav.\/}
{\bf20}, 1249-1262 (1988).

\refis{Gibb}P.E. Gibbs, ``The small-scale structure of spacetime: A
bibliographical review,"\cut hep-th/9506171.

\refis{Grom}M. Gromov, {\sl Structures m\'etriques pour les vari\'et\'es
riemanniennes\/} (Paris: Cedic/F Nathan 1981).

\refis{Gupt}S. Gupta, ``Causality in spin foam models," {\sl Phys. Rev.} D
{\bf61} (2000) 064014, and gr-qc/9908018.

\refis{Hawk}S.W. Hawking, ``Space-time foam" {\sl Nucl. Phys.\/} {\bf B144},
349-362 (1978).

\refis{HaEl}S.W. Hawking and G.F.R. Ellis, {\sl The Large Scale Structure of
Space-Time\/} (Cambridge University Press 1973).

\refis{HKMC}S.W. Hawking, A.R. King and P.J. McCarthy, ``A new topology for
curved space-time which incorporates the causal, differential and conformal
structures," {\sl J. Math. Phys.} {\bf17}, 174-181 (1976).

\refis{Iwas}J. Iwasaki, ``A surface theoretic model of quantum gravity,"
gr-qc/9903112.

\refis{Kemp}A. Kempf, ``On the structure of spacetime at the Planck scale," 
hep-th/9810215.

\refis{KlRo}D.J. Kleitman and B.L. Rothschild, ``Asymptotic enumeration of
partial orders on a finite set," {\sl Trans. Am. Math. Soc.} {\bf205},
205-220 (1975).

\refis{Mala}D. Malament, ``The class of continuous timelike curves determines
the topology of spacetime," {\sl J. Math. Phys.} {\bf18}, 1399-1404 (1977).

\refis{MaSm}F. Markopoulou and L. Smolin, ``Quantum geometry with intrinsic
local causality," {\sl Phys. Rev.} D {\bf58}: 084032 (1998), and
gr-qc/9712067.

\refis{Meye}D.A. Meyer, {\sl The Dimension of Causal Sets}, PhD thesis, MIT
1988.

\refis{MeSo}D.A. Meyer and R.D. Sorkin, unpublished work.

\refis{Myrh}J. Myrheim, ``Statistical geometry," 1978 preprint, Ref. TH
2538-CERN, unpublished.

\refis{OvWe}J.M. Overduin and P.S. Wesson, ``Kaluza-Klein gravity," {\sl
Phys. Rep.} {\bf283}, 303-380 (1997), and gr-qc/9805018.

\refis{ReRo}M.P. Reisenberger and C. Rovelli, ``Sum over surfaces form of loop
quantum gravity," {\sl Phys. Rev.\/} D {\bf56}, 3490-3508 (1997), and
gr-qc/9612035.

\refis{Seri}M. Seriu, ``The spectral representation of the spacetime
structure: The `distance' between universes with different topologies," {\sl
Phys. Rev.\/} D {\bf53}, 6902-6920 (1996), and gr-qc/9603002;
\new\dash ``Space of spaces as a metric space," {\sl Commun. Math. Phys.}
{\bf209}, 393-405 (2000), and gr-qc/9908078.

\refis{Ser2}M. Seriu, ``Spectral representation and the averaging problem in
cosmology,"\cut gr-qc/0001014, to appear in {\sl Gen. Rel. Grav.}

\refis{Sork}R.D. Sorkin, ``A specimen of theory construction from quantum
gravity," in {\sl The Creation of Ideas in Physics}, J Leplin, ed (Kluwer
1995), and gr-qc/9511063.

\refis{Szab}L.B. Szabados, ``Causal measurability in chronological spaces,"
{\sl Gen. Rel. Grav.} {\bf19}, 1091-1100 (1987).

\refis{Hoof}G. 't Hooft, ``Quantum gravity: a fundamental problem and some
radical ideas," in {\sl Recent Developments in Gravitation. Carg\`ese 1978},
M. L\'evy and S. Deser, eds. (Plenum 1979).

\refis{Wald}R.M. Wald, {\sl General Relativity\/} (University of Chicago Press
1984).

\refis{Whee}J.A. Wheeler, ``On the nature of quantum geometrodynamics,"
{\sl Ann. Phys.} {\bf2}, 604-614 (1957);
\new\dash {\sl Geometrodynamics\/} (Academic Press 1962), pp 71--83.

\refis{Woot}W.K. Wootters, ``Statistical distances in Hilbert spaces," {\sl
Phys. Rev.} D {\bf23}, 357-362 (1981).

\refis{Zala}R. Zalaletdinov, ``Approximate symmetries in general relativity,"
gr-qc/9912021.

\endreferences
\endit
\baselineskip18pt
\centerline{\bf Figure Captions}
\bigskip
\noindent{\bf Figure 1:} The geometry $G_\gamma$. The drawing shows the case
$b<a$, or $\gamma<1$, with a sprinkled point ${\bf x}_1$ and its future light
cone. For this particular point, $V(I^+({\bf x}_1)) = (b-t)^2 - \half\,
[(b-t)-(a+x)]^2$, where the area of the small triangle with dashed edges in
the upper right hand corner must be subtracted, since $a-x < b-t$.
\bigskip
\noindent{\bf Figure 2:} The geometry $G'_\gamma$. The drawing shows
the case $b>a/2$, or $\gamma>\half$, with a sprinkled point ${\bf x}_1$ and
its future light cone. For this particular point, $V(I^+({\bf x_1})) =
(b-t)^2 - (b-t-a/2)^2$, where the area of the two small triangles with dashed
edges must be subtracted, since $b-t > a/2$. Any two points whose $x$
coordinates differ by $a$ are identified; in particular, the two outer
vertical dashed lines are to be identified with each other.
\bigskip
\noindent{\bf Figure 3:} Plot of $d_2(G_\gamma,G'_\gamma)$ as a function
of $\gamma$, for $0<\gamma<5$; for $\gamma>5$, the function decreases
monotonically, and approaches zero as $\gamma\to\infty$.

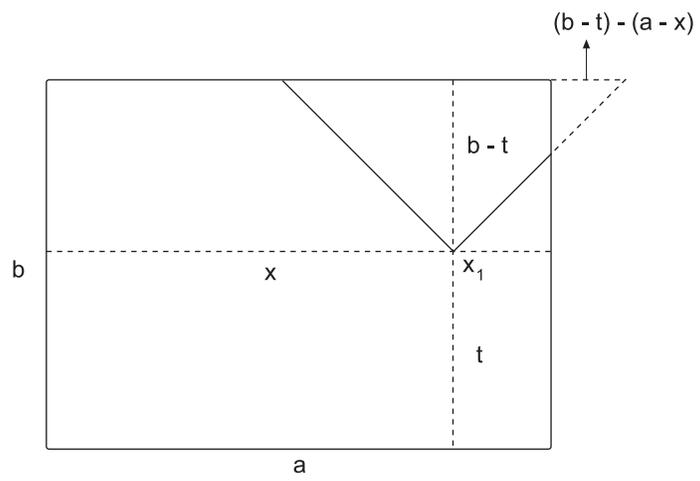
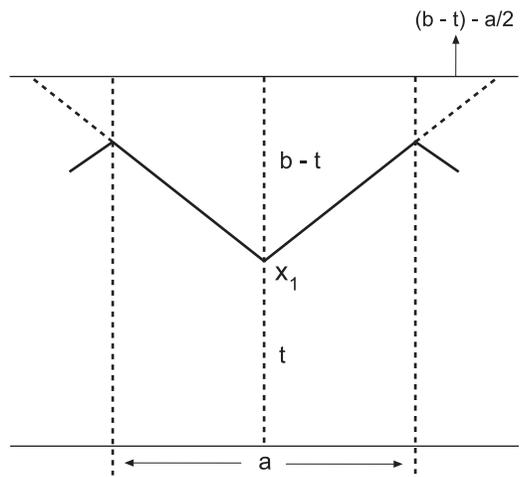
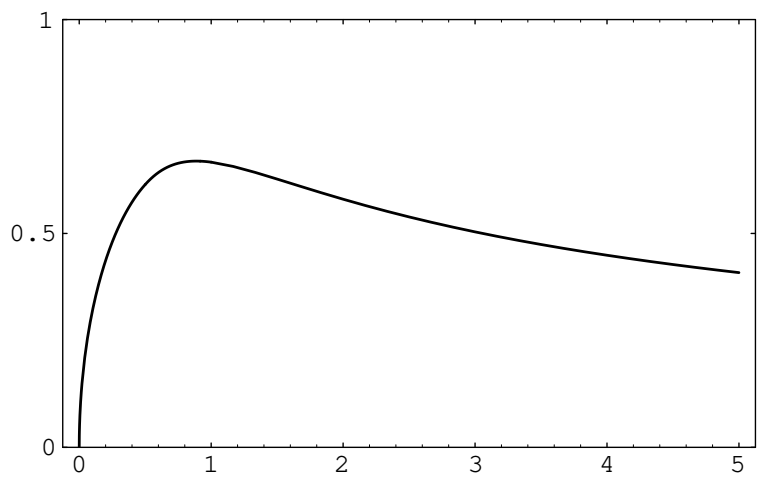
\end